\documentclass[12pt]{iopart}

\usepackage{cite}

\usepackage[english]{babel}
\usepackage{latexsym}
\usepackage{graphics,psfrag}
\usepackage{epsfig}

\begin{document}

\title{Many-body Anderson localization in one dimensional systems}

\author{Dominique Delande}
\address{Laboratoire Kastler Brossel, UPMC-Paris6, ENS, CNRS; 4 Place Jussieu, F-75005 Paris, France}

\author{Krzysztof Sacha} 
\address{
Instytut Fizyki imienia Mariana Smoluchowskiego, 
Uniwersytet Jagiello\'nski, ul.~Reymonta 4, PL-30-059 Krak\'ow, Poland}
\address{
Mark Kac Complex Systems Research Center, 
Uniwersytet Jagiello\'nski, ul.~Reymonta 4, PL-30-059 Krak\'ow, Poland}

\author{Marcin P\l odzie\'n} 
\address{
Instytut Fizyki imienia Mariana Smoluchowskiego, 
Uniwersytet Jagiello\'nski, ul.~Reymonta 4, PL-30-059 Krak\'ow, Poland}

\author{Sanat K. Avazbaev} 
\address{Laboratoire Kastler Brossel, UPMC-Paris6, ENS, CNRS; 4 Place Jussieu, F-75005 Paris, France}
\address{ARC Centre for Antimatter-Matter Studies, Curtin University of Technology, GPO Box U1987, Perth, Western Australia 6845, Australia}

\author{Jakub Zakrzewski} 
\address{
Instytut Fizyki imienia Mariana Smoluchowskiego, 
Uniwersytet Jagiello\'nski, ul.~Reymonta 4, PL-30-059 Krak\'ow, Poland}
\address{
Mark Kac Complex Systems Research Center, 
Uniwersytet Jagiello\'nski, ul.~Reymonta 4, PL-30-059 Krak\'ow, Poland}

\pacs{72.15.Rn, 03.75.Lm, 37.25.+k, 71.23.An}

\begin{abstract}
We show, using quasi-exact numerical simulations, that 
Anderson localization in a disordered one-dimensional potential survives in the presence of attractive interaction
between particles. The localization length of the particles center of mass - computed
analytically for weak disorder - is in good agreement with the quasi-exact numerical observations
using the Time Evolving Block Decimation algorithm. Our approach allows for simulation of the entire 
experiment including the final measurement of all atom positions.
\end{abstract}

\maketitle

\section{Introduction}

Anderson localization (AL) has been widely investigated in the last 50 years~\cite{Anderson1958,Lagendijk2009}. 
The possibility of observing directly localization of the wavefunction in cold atomic gases
lead to a recent revival of the interest in localization properties
in general, and in AL in particular. AL is characterized by 
the inhibition of transport in a quantum system, whose classical counterpart behaves diffusively. 
It is accompanied by an exponential localization of eigenstates in the configuration
space, $|\psi(r)|^2 \propto \exp(-|r|/L),$ where $L$ is the localization length. 
AL is due to the interference between various multiple scattering 
paths which favors the return of the particle to its initial position and thus decreases its probability
to travel a long distance. As the geometry of these paths  depends
on the system dimension, so  AL features depend on the dimension too.
For  one-dimensional (1D) systems, AL is a generic single particle behavior even for very small disorder
when the particle ``flies'' above the potential fluctuations. 

A fundamental question is to understand how interaction between particles affects AL. 
Presently, there is no consensus on the possible existence and properties of many-body localization. Some
results suggest that AL survives for few-body systems, although with a modified localization length~\cite{Shepelyansky1994};
studies of cold bosonic systems in the mean-field regime show that AL is destroyed and replaced by a sub-diffusive
behavior~\cite{Pikovsky2008,Flach2011}, but the validity of the mean-field approximation
at long times is questionable. There are even predictions that AL survives at finite temperature in the thermodynamic
limit,
in the presence of interactions~\cite{Aleiner2010}. In this paper, we show, using a specific example,
that 1D AL survives in the presence of attractive interactions and is even a rather robust phenomenon. 
The novelty of our approach is that it uses a quasi-exact numerical scheme to solve the full many-body problem
in the presence of disorder. Here, quasi-exact means that all numerical errors can be controlled and reduced
below an arbitrary value, just at the cost of increased computational resources. The big advantage of this approach
is not to rely on neglecting \emph{a priori} any physical process.

Atomic matter waves have several advantages that made possible an unambiguous demonstration of single particle AL in 
1D \cite{Billy2008,Roati2008}: atom-atom interaction can be reduced
either by diluting the atomic gas or using Feshbach resonances, ensuring a very long
coherence time of the atomic matter wave; the spatial and temporal orders
of magnitudes are very convenient allowing a direct spatio-temporal visualization of the dynamics;
all microscopic ingredients are well controlled; a disordered potential can be created by using the
effective potential induced by a far detuned optical speckle. 

\section{The model and its solution}
We consider $N$ identical bosonic atoms in a 1D system, in the regime of attractive two-body interactions. 
We assume the dilute regime where the atom-atom average distance is larger than the scattering length
and take the low-energy limit where the interaction can be modeled by a (negative) Dirac-delta potential. 
The many-body Hamiltonian can be written, using the standard second quantization formalism
(assuming unit mass for the particles and
taking $\hbar=1$):
\begin{eqnarray}
\hat{H} = \int{dz\ \hat{\psi}^{\dagger}(z) \left[ -\frac{1}{2}\ \frac{\partial^2}{\partial z^2} + V(z) \right] \hat{\psi}(z)} \nonumber \\
+ \frac{g}{2} \int{dz\ \hat{\psi}^{\dagger}(z)\ \hat{\psi}^{\dagger}(z)\ \hat{\psi}(z)\ \hat{\psi}(z)}
 \label{eq:h}
\end{eqnarray}
where $g<0$ is the strength of the atom-atom interaction and $V(z)$ an external potential.

For large $N$, in the absence of an external potential, the ground state of this system is 
described - within the mean field approach - as a bright soliton,
 a composite particle  with two external degrees of freedom: an irrelevant
phase $\theta$, and a classical parameter: the position $q$ of the center of mass.
The particle density -- normalized to the number $N$ of particles -- 
is given by $|\phi_0(z-q)|^2$ where 
\begin{equation}
 \phi_0(z) = \sqrt{\frac{N}{2\xi}} \frac{{\mathrm e}^{-i\theta}}{\cosh{z/\xi}}.
\label{eq:shape}
\end{equation}
$\xi=-\frac{2}{Ng}$ is the characteristic size of the soliton. 
The associated chemical potential is $\mu=-N^2g^2/8.$ 

This mean-field approach does not describe properly the many-body ground state of the system. 
Indeed, in the absence of an external potential, 
the many-body ground state is known exactly, thanks to the Bethe ansatz~\cite{McGuire1964} which predicts, e.g., uniform atomic density. 
The source of discrepancy lies in a classical treatment of the center of mass $q$
of the system. 
In a proper description, $q$ must be thought of as the quantum position operator of the soliton, the composite particle formed by the $N$ particles. 
In the presence of an external potential, it is possible to construct an effective one-body (EOB) Hamiltonian describing the $q$ dynamics 
quantum mechanically \cite{Weiss09,Sacha2009a,Sacha2009b,Mueller2011}. 
Assuming a fixed soliton shape, the EOB Hamiltonian is:
\begin{equation}
 H_q = \frac{p_q^2}{2N} + \int{\ dz \ |\phi_0(z-q)|^2\ V(z)},
\label{eobpot}
\end{equation}
where $p_q$ is the momentum conjugate to $q$ \cite{Weiss09,Sacha2009a,Sacha2009b,Mueller2011}. 
It describes a composite particle with mass $N$ evolving in a potential that is the convolution of the bare potential with the soliton envelope.
The key point of the EOB approach is that the internal degrees of freedom of the soliton are hidden in the reduction of the 
many-body wavefunction to a single one-body wavefunction $\varphi(q,t)$ describing the evolution of the soliton center of mass.  
This is possible because the internal degrees of freedom of the bright soliton are gaped, 
with an energy gap equal to $-\mu=N^2g^2/8,$ so that a weak external perturbation cannot populate internal 
excited states of the bright soliton, in contrast with the dark soliton case~\cite{Mochol2012}.

Because the EOB Hamiltonian~(\ref{eobpot}) describes a one-dimensional system exposed to a disordered potential, it displays Anderson localization.
Within the EOB approximation, the soliton center of mass is localized with a 
localization length depending on the energy. As an example, we will use - as in real experiments~\cite{Billy2008} - the disorder created by a light speckle 
shone on a cold atomic gas. The localization length of the EOB model 
has been calculated (in the weak disorder limit) 
in~\cite{Sacha2009a,Sacha2009b,Mueller2011}:
\begin{equation}
 \frac{1}{L_N(k)} = \frac{N^4 \xi^2 \pi^3 \sigma_0 V_0^2 (1-k\sigma_0)}{\sinh^2 \pi k\xi}\ \Theta(1-k\sigma_0)
\label{eq:locN}
\end{equation}
where $k$ is the wave-vector of the soliton, $V_0$ the r.m.s amplitude of the disordered potential, $\sigma_0$ its correlation length
of the speckle and $\Theta$ the Heaviside function.

What is the validity of the EOB theory at long time? The answer is far from obvious. Any many-body effect not taken into account within the EOB, could break the
reduction of the problem to an EOB wavefunction evolving under the effective Hamiltonian~(\ref{eobpot}). Especially, it could easily 
spoil the phase coherence of the EOB wavefunction and consequently Anderson localization. It is the goal of this paper to perform a quasi-exact many-body
numerical test of the EOB approach. In order to be as close as possible to a realistic experiment~\cite{Billy2008}, 
we follow the temporal evolution of an initially localized many-body wavepacket.
In a first step, we compute the ground state of $N$ interacting particles in the presence of an harmonic trap, but
without disorder: this produces a bright soliton localized near the trap center. In a second step, the harmonic trap is abruptly switched off and
the disordered speckle potential abruptly switched on, leaving the many-body system expand in the presence of disorder, and eventually
localize thanks to many-body AL.

In the presence of an external potential, the many-body problem cannot be solved exactly. One must rely on quasi-exact numerical approaches.
A convenient way is  to
discretize the continuous Hamiltonian, Eq.~(\ref{eq:h}),  over a discrete
lattice~\cite{Schmidt2007,Glick2012}. 
The discretization of space to a chain of sites 
located at equally spaced positions $z_l=l\delta$ together with a
 3-point discretization of the Laplace operator allows us to write the Hamiltonian in a tight-binding Bose-Hubbard form 
\cite{Schmidt2007}:
\begin{equation}
H=\sum_l{\left[-J(a_l^{\dagger}a_{l+1} + h.c.) + \frac{U}{2} a_l^{\dagger}a_l^{\dagger}a_{l}a_{l} + V_l\ a_l^{\dagger}a_{l}\right]}
\label{eq:h_discrete} 
\end{equation}
with $J=\frac{1}{2\delta^2}$, $U=\frac{g}{\delta}$ and $V_l=V(z_l)$ (an additional trivial constant term $1/\delta^2$ has 
been dropped).

For numerical purposes the infinite space is restricted to the $[-K\delta,K\delta]$ interval leading to a 1D chain of $M=2K+1$
discrete sites.  $N$ identical bosons are distributed on these $M$ sites.
A basis of the total Hilbert space can be built using the direct product of Fock states on each site 
$|i_1,i_2\ldots i_M\rangle$ with the constraint that the sum of 
occupation numbers $\sum_{l=1..M}i_l$ is equal to $N.$
The size of this basis increases exponentially with the system size, making its use unpractical for large many-body problems. 
Instead, we use a variational set of Matrix Product States (MPSs). An MPS is a state which can be written
as:
 \begin{equation}
|\psi\rangle\! = \!\!\!\!\!\!\sum\limits_{\alpha_1,\ldots,\alpha_M;\  i_1,\ldots,i_M}\!\!\!\!\!
\Gamma_{1\alpha_1}^{[1]
,i_1}\lambda^{[1]}_{\alpha_1}\Gamma^{[2],i_2}_{\alpha_1\alpha_2}\ldots\Gamma^{[M],i_M}_{\alpha_{n-1}1} |i_1,\dots,i_M\rangle
\label{eq:MPS}
\end{equation}
where $\Gamma^{[l],i_l}$ ($\lambda^{[l]}$) are site (bond) dependent matrices (vectors).
To describe exactly a generic state in terms of MPS, a large number (exponentially increasing with $M$)
of $\alpha_l$ values is needed. However, typical low energy states  are only slightly entangled so 
that $\lambda^{[l]}_{\alpha_l=1,2\ldots}$ are rapidly
decaying numbers, which allows for introduction of a  cutoff $\chi$ in the sum over Greek indices above,
resulting in tractable numerical computations~\cite{Vidal2003}. 
For a ground state protected by a gap, the area theorem \cite{Eisert2010} ensures that an efficient MPS representation exists. 
In the physical situation discussed by us, the area law has no direct applicability.

The  $i_1,..i_M$ indices are in principle restricted to the $[0,N]$ interval.
In practice, since it is highly unlikely that all the bosons occupy a single site of the system, 
we lower the cutoff in the sums assuming some $N_{\textrm{max}}<N$. While the maximum average occupation number (at sites
near the center of mass of the soliton) is $N\delta/2\xi$=2.5, 
we found surprisingly that a relatively high $N_{\textrm{max}}=14$ is needed for convergent results, while the $\chi$ value may be
kept relatively low.

The ground state as well as the dynamics may be quasi-exactly 
studied using the Time Evolving Block Decimation (TEBD) algorithm~\cite{Vidal2003,Vidal2004}, 
essentially equivalent to the
time-dependent Density Matrix Renormalization Group approach~\cite{White2004,Schollwock2011}.  
 The TEBD  algorithm describes how the $\Gamma^{[l]}$ and $\lambda^{[l]}$ 
evolve in time under the
influence of an Hamiltonian containing simple terms local on each site as well as hoping terms
of the type $a_l a_{l+1}^{\dagger}$ which transfer one particle from site $l$ to site $l+1.$
A maximum of $N=$25 particles
could be included in our calculations; similar results are obtained for 10 particles. 

We use the soliton size $\xi$ as the unit of length; consequently 
the time unit is $\xi^2.$  We use the EOB as a guide to choose the parameters of the 
many-body numerical experiment 
For example, the trap must be shallow enough not to distort the soliton shape (\ref{eq:shape}), but still strong enough to confine
its center of mass $q$ over a distance only slightly larger than its size. We chose $\Delta q = \sqrt{\frac{8}{5}} \xi;$
The frequency, $\omega$, of the trapping harmonic potential $\omega^2z^2/2$ is thus such that
$N\omega= 5/8\xi^2,$ i.e. $\omega=0.025/\xi^2.$ 
In order for the
localization length to be reasonably short, we choose
the strength of the external potential comparable to the initial average energy of the soliton
$\omega/4$, that is $V_0=2.5\times10^{-4}.$ We also choose the
correlation length of the speckle potential $\sigma_0=0.4\xi$ to be significantly shorter than the soliton size, so
that the EOB potential in Eq.~(\ref{eobpot}) is 
free of the peculiarities of the speckle potential~\cite{Lugan2009}.

\begin{figure}
\begin{center}
\resizebox{0.9\columnwidth}{!}{\includegraphics{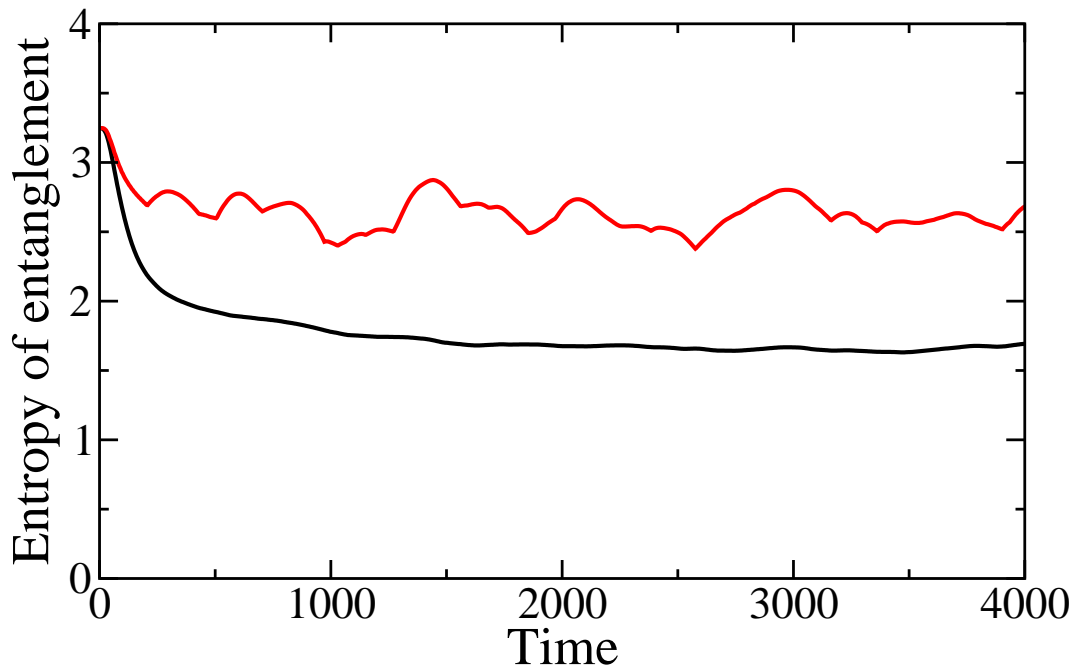}} 
\caption{Time evolution of the entropy of entanglement, Eq.~(\ref{entropydef}), for an initially localized bright soliton 
expanding in a disordered speckle potential, obtained using the many-body TEBD algorithm. No significant growth of the entropy is observed. The red curve is the largest entropy among 96 different realizations of disorder while the black curve corresponds to the average over these realizations. Parameters are given in the text.}
\label{entrop}
\end{center}
\end{figure}

Several sources of errors exist in the TEBD algorithm and must be controlled. The first one is due to the spatial discretization.
Getting accurate results requires the discretization unit, $\delta$, to be much smaller than both the soliton
size $\xi$ and the de Broglie wavelength $2\pi/k$, where $k$ is the typical wave-vector contained in the initial wavepacket for the 
center of mass in the EOB description. We use $\delta=\xi/5;$ a twice smaller step produces
slightly different quantitative results, but the difference is practically invisible on the scale of the figures shown in our work. 
In order to avoid reflections from the boundaries, the number of lattice points must be sufficiently large; 
1921 points are used, but only the central 1201 ones are shown in the plots.
A second source of error is the temporal discretization of the evolution operator. We use the standard Trotter
expansion~\cite{Vidal2003}, whose error can be controlled by varying the time step ($\delta t=0.008\xi^2$ is used). 
A third source of error is the truncation of the MPS at each step. This error is monitored through the so-called
``discarded weight'', that is the weight of the components which have to be discarded from the time evolved many-body state
to keep it in MPS form with a fixed parameter $\chi$ -- the number of bonds between sites. This can be a serious
problem when the entropy of entanglement grows as a function of time. It may even prevent calculation to be ran beyond some rather short time,
especially when the system is significantly excited above the ground state~\cite{Daley2004}.

The entropy of entanglement is defined as the supremium over all possible bipartitions of the system. Explicitly we compute
\begin{equation}
S=\sup_l  S_l =\sup_l \left[ - \sum_{\alpha}{ (\lambda^{[l]}_{\alpha})^2 \ln(\lambda^{[l]}_{\alpha})^2 }\right]
\label {entropydef}
\end{equation}
with $l$ running over all bonds. Typically, the maximum is reached for a link close to the center of the system but it can depend on the disorder and fluctuate in time. Quite surprizingly, we have not observed any significant growth of the entropy of entanglement when AL sets in, see Figure~\ref{entrop}, a result quite opposite to that observed in~\cite{Bardarson2012}. This may be attributed to the fact that 
 the energy of our many-body state is quite small (see the discussion above).
 The converged calculations reported here use $N_{\textrm{max}}$=14, $\chi$=30 yielding the internal Hilbert space
dimension per site being 450. The results has 
been compared with those with lower $N_{\textrm{max}}$ as well as those for $\chi$=40 (for shorter times) to check that
the results presented are fully converged. All in all we were able
to run the fully controlled  calculation up to time $t=4000$.

\begin{figure}
\begin{center}
\resizebox{0.95\columnwidth}{!}{\includegraphics{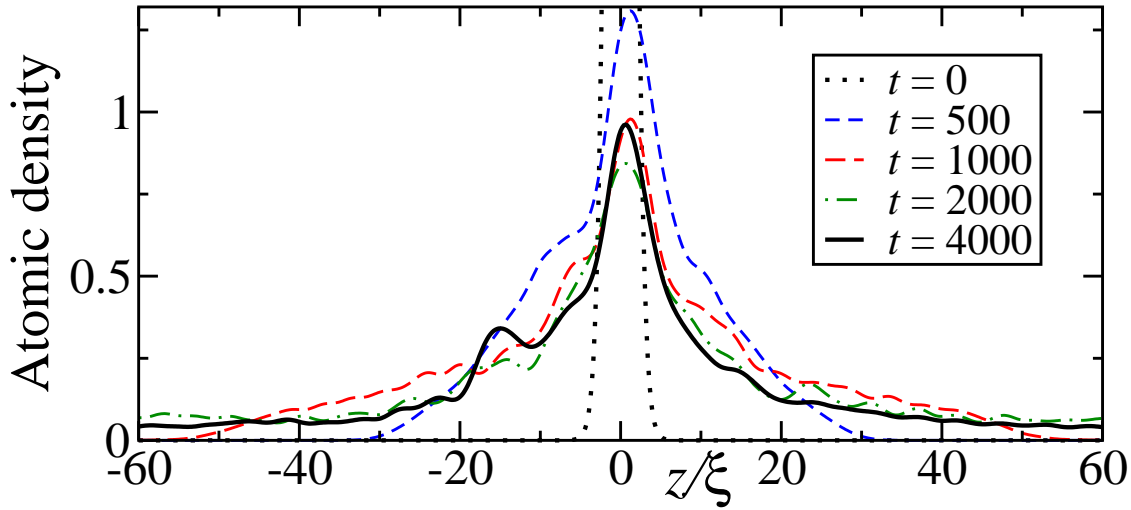}} 
\caption{Spatial atomic density of an initially localized bright soliton, under the influence
of a disordered speckle potential, using the quasi-exact many-body TEBD algorithm, at various times. 
After an initial expansion, the atomic density freezes at long time, a signature of many-body Anderson localization. 
Parameters are given in the text.}
\label{fig:density}
\end{center}
\end{figure}

Figure~\ref{fig:density} shows the particle density in configuration space obtained at increasing
times averaged over 96 realizations
of the disorder. 
In the absence of disordered potential, the initial wavepacket is expected to spread. 
The EOB physics gives 
the characteristic time for this spreading, $t=1/\omega=40,$ much shorter than the time scale in the figure.
At $t$=500 and 1000, one clearly sees that the central part of the wavepacket is already more or less localized while
 the ballistic front for $|z|/\xi>20$ ($|z|/\xi>40$ for $t$=1000) has not yet been scattered and keeps a Gaussian shape similarly as in the EOB description. This corresponds
to the wavepacket components with the highest energy and consequently the longest localization length. 
AL has already setup at $t$=2000 and does no longer evolve further, compare with $t$=4000.
Therefore, Fig.~\ref{fig:density} provides an evidence for many-body AL taking place in a 
quasi-exact full many-body numerical simulation. At the final time (100 times the characteristic spreading time),
we do not observe any indication that AL could be destroyed.

\begin{figure}
\begin{center}
\resizebox{0.6\columnwidth}{!}{\includegraphics{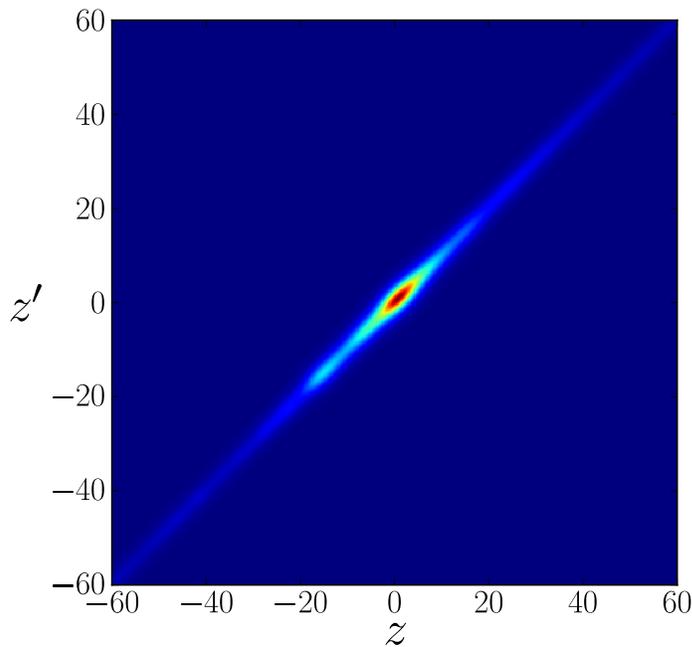}} 
\caption{ 
One-body density matrix
of the many-body state at $t=4000$ (see Fig.~\ref{fig:density}) 
in configuration space. It is strongly concentrated along the diagonal $z=z'$ with a transverse
width of the order of the soliton size $\xi.$ This is a direct proof that the many-body
system can be described by a compact composite particle, a bright soliton, whose center of mass is widely spread.
}
\label{fig:one-body}
\end{center}
\end{figure}

The description of the final state as a MPS makes it possible to compute easily more complicated quantities, such as
correlation functions. The simplest one is the one-body density matrix 
$\langle \psi(z) \psi^{\dagger}(z')\rangle,$ shown in Fig.~\ref{fig:one-body}.
It clearly displays extremely strong correlations between positions $z$ and $z',$ reinforcing the
observation that AL probably survives far beyond $t=$4000.
The interpretation is simple in terms of bright soliton: all atoms are grouped in a soliton of size $\xi,$
but the center of mass of the soliton itself is widely spread. This has an important consequence: the largest
eigenvalue of the one-body density matrix -- a value often used as a quantitative criterion for Bose-Einstein
condensation \cite{Pethick} -- is here 0.14, much smaller than unity; in contrast, the value at $t=0$ is 0.84.
Thus, while the initial state can be considered as a true condensate, 
the temporal dynamics destroys condensation; any description of our many-body system using the mean field theory via 
the Gross-Pitaevskii equation (which by construction describes a 100\% condensate) \emph{must} fail. 
In other words, our many-body AL
is necessarily beyond the mean field description.

\section{Simulation of a measurement}

\begin{figure}
\begin{center}
\resizebox{0.95\columnwidth}{!}{\includegraphics{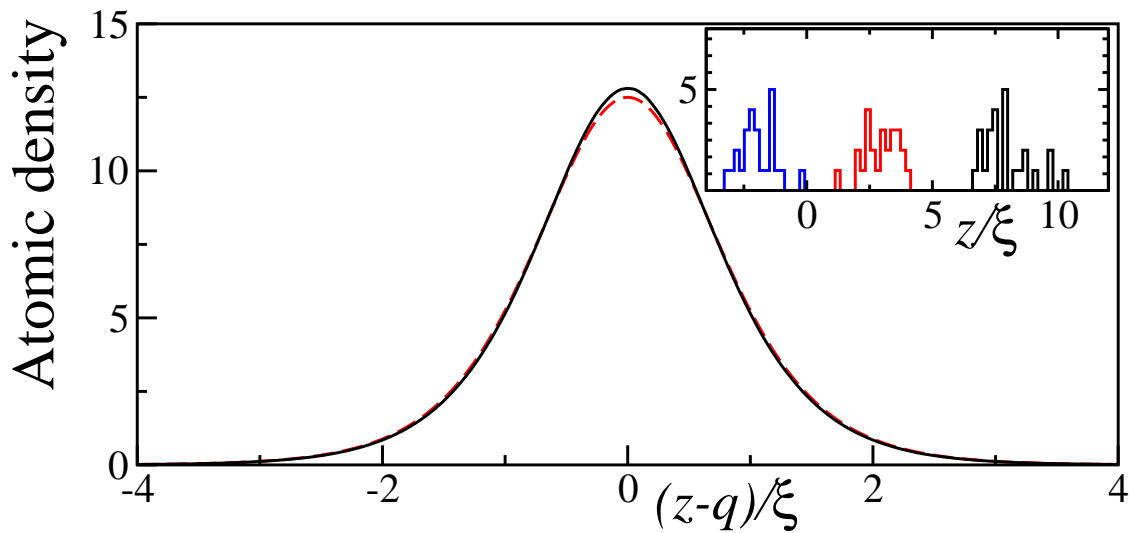}} 
\caption{ Atomic density measured with respect to the center of mass of the $N$ particles,
at time $t=4000,$ after Anderson localization has set in. A large number of
individual ``measurements'' of the particle positions -- three examples
are shown in the inset -- are performed for various disorder realizations, and
the resulting average is displayed in the main panel.
The agreement between the full many-body
calculation (solid line) and the
prediction of Eq.~(\ref{eq:shape}) for the soliton shape (dashed line) is excellent. 
}
\label{fig:center-of-mass}
\end{center}
\end{figure} 

Being able to write the many-body state as a MPS has considerable further advantages, especially if it is -- like in our calculations -- 
in the so-called canonical form~\cite{Schollwock2011}. For example, expectation values of local operators
such as $a_l^{\dagger}a_l$ or $a_l^{\dagger}a_{l+1}$ involves only simple contractions on the local $\Gamma^{[l]}$ tensors
and $\lambda^{[l]}$ vectors. It makes it also possible to mimic the measurement process of particle positions as follows.
The reduced density matrix $\rho^{[l]}$ on site $l$ is easily constructed by contracting the $\Gamma^{[l]}$ tensor with the
neighboring
$\lambda^{[l-1]}$ and  $\lambda^{[l]}$ vectors:
\begin{equation}
 \rho^{[l]}_{i,j} = \sum_{\alpha_{l-1},\alpha_l}{[\lambda^{[l-1]}_{\alpha_{l-1}}]^2\ \Gamma^{[l],i}_{\alpha_{l-1}\alpha_l}
\ [\Gamma^{[l],j}_{\alpha_{l-1}\alpha_l}]^* \ 
[\lambda^{[l]}_{\alpha_{l}}]^2}
\end{equation}

We then chose randomly the number of particles ``measured'' on site $l$ following the statistical populations,
diagonal elements of the on-site reduced density matrix. Once a given occupation number $i$ is chosen,
we project the MPS state onto the subspace with $i_l=i$ and normalize it. This involves only
simple contractions on the local $\Gamma$ tensors and $\lambda$ vectors, producing another MPS. The process
can be iterated on all sites, and is particularly simple if sites are scanned consecutively starting from
one edge and propagating toward the other edge. It is simple to prove that the probability distribution
of the measurements is independent of the order used for scanning the various sites.

An individual ``measurement'' produces a single set of occupation numbers $(i_1,i_2...i_M)$ (whose sum is of course $N$)
whose probability is exactly $|\langle i_1,i_2...i_M|\psi\rangle|^2.$ By performing a series of ``measurements'', we can
sample interesting physical quantities, such as the position of the soliton center of mass, 
$\langle q\rangle = \sum_l{l\delta i_l}/N,$ and the particle density with respect to this center of mass.
Note that these quantities are hard to measure by other means as they involve correlation functions
of high order (typically up to $N$)~\cite{Dziarmaga2010,Mishmash2010}.

Examples of our procedure are given in the inset of Fig.~\ref{fig:center-of-mass}. 
In this way, we extract both the position of the center
of mass of the soliton
and the atomic density relative to the center of mass. 
The later quantity is shown in Fig.~\ref{fig:center-of-mass} for time $t=4000$ in comparison
with the analytic prediction, Eq.~(\ref{eq:shape}). The agreement is excellent,
showing that the internal structure of the soliton is fully preserved for a long time,
even after AL has set in. The small difference is a $1/N$ finite size effect.

\section{Comparison with effective one body approach}

The EOB theory  is able to quantitatively predict the long time limit
for the spatial density probability of the soliton center of mass, see the detailed derivation and calculations
in~\cite{Sacha2009a}. Initially, the center of mass is in the ground state of the harmonic trap (a Gaussian wavepacket)
that, after the trap is switched off, expands over a range of energies, each energy component being characterized by its own
localization length.
Each component displays approximate  exponential localization in the long
time limit (in a 1D system, Anderson localized eigenstates do not strictly decay exponentially,
see e.g. \cite{gogolin,muellerdel}). Their superposition displays approximate algebraic localization
at long distance, as discussed in~\cite{Sacha2009a}. 

\begin{figure}
\begin{center}
\resizebox{0.9\columnwidth}{!}{\includegraphics{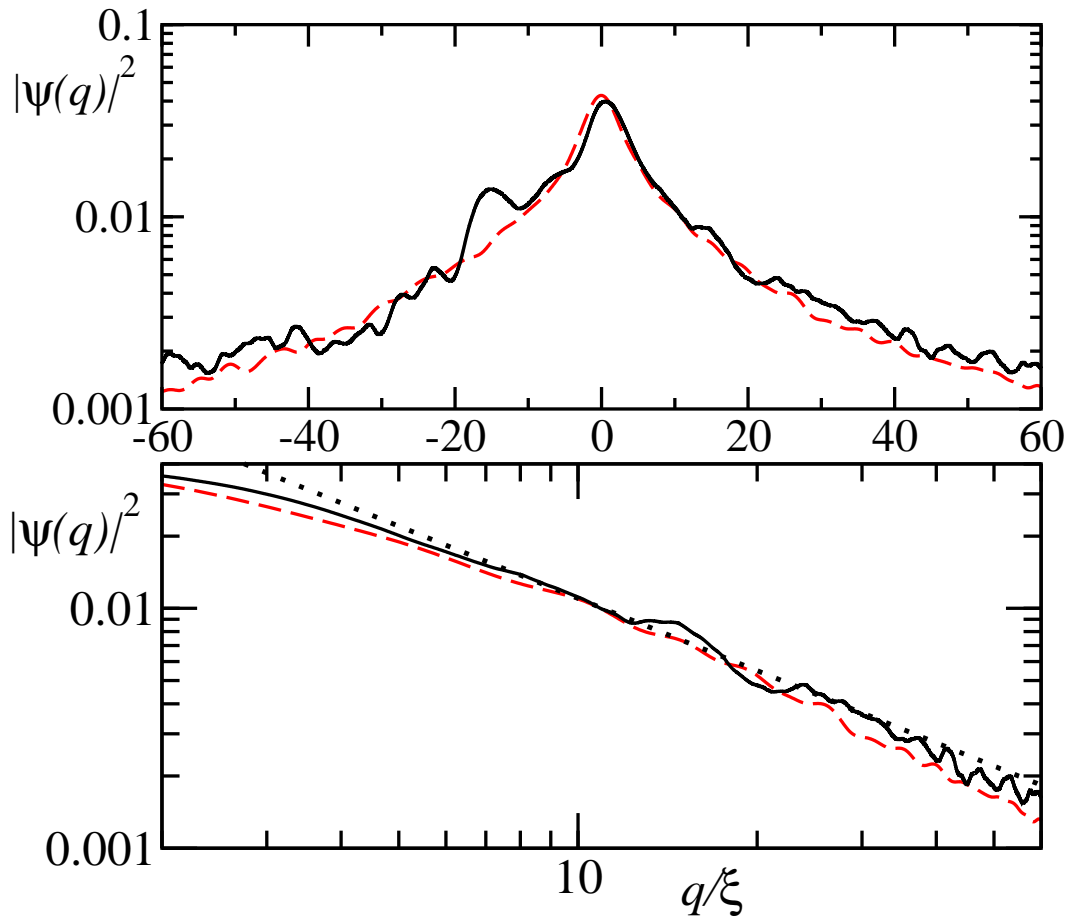}} 
\caption{ Probability density for the center of mass of the soliton at time $t=4000$,
numerically computed using the many-body quasi-exact dynamics (solid line, averaged over 96 disorder
realizations) and the effective one-body theory (dashed line, averaged over 10000 realizations).
The good agreement shows that the many-body problem actually displays Anderson localization, as
predicted by a simple theory for a composite particle. The dotted line in the lower panel (double logarithmic
plot) is the $1/|q|$
approximate prediction of the effective one-body theory \cite{Sacha2009a}.}
\label{fig:comparison}
\end{center}
\end{figure}

In Fig.~\ref{fig:comparison}, we show 
comparison between the full many-body calculation and  the EOB corresponding numerical simulation with Hamiltonian~(\ref{eobpot}). Note
that the many-body result is here plotted for the center of mass position, which can slightly
differ from the atomic density; the latter, in the EOB approach, is the convolution of the former by the soliton shape.
At the scale of the figure, the soliton is extremely narrow so that the result of the convolution is almost equal the center of mass density,
compare with Fig.~\ref{fig:density}. The agreement between the many-body and the EOB calculations is clearly excellent. In Fig.~\ref{fig:comparison},
we also show the $1/|q|$ leading behavior predicted by the EOB theory, Eq.~(13) of Ref.~\cite{Sacha2009a}.
It predicts quite well the observed behavior but does not aim at being quantitative, because of
the existence of a sub-leading logarithmic term. Namely,
at very large distance, the exponential term $\exp[-\beta \ln^2(\gamma|q|)]$, where $\gamma$ and $\beta$ are constants, present in Eq. (13) 
of Ref.~\cite{Sacha2009a} becomes important, leading to a faster decrease of the distribution and eventually to a 
finite rms displacement $\langle q^2\rangle$ of the soliton.
Note also that  the formula, Eq.~(13) of Ref.~\cite{Sacha2009a}, assumes a weak
disorder (Born approximation), an assumption not fully satisfied here.

Finally, we compare the localization length of the center of mass of attractively interacting particles 
with the localization length of a single particle in the same disordered potential. 
A meaningful comparison must be performed for the
same total energy per particle, or equivalently for the same wave-vector per particle; this thus corresponds to a wavector $N$ times larger
for the soliton, composed of $N$ individual particles. 
Within the EOB approach, the ratio is, for $k\sigma_0<1$ and weak disorder:
\begin{equation}
 \frac{L_{1}(k/N)}{L_{N}(k)} = N^2 \left[\frac{\pi k \xi}{\sinh \pi k \xi}\right]^2\ \frac{1-k\sigma_0}{1-k\sigma_0/N}
\end{equation}
The physical interpretation is simple and interesting. The $N^2$ factor strongly favors localization of the soliton and reflects
the collective behavior of the $N$ attractive bosons when placed in the disordered potential. The second factor -- and to a lesser extent, the third one -- 
is smaller than unity and favors delocalization of the soliton. It reflects the fact that the center of mass of the soliton does not feel the raw potential,
but rather its convolution with the soliton shape, see eq.~(\ref{eobpot}); being smoother, the convoluted disordered scatters less efficiently than the raw one,
leading to an increase of the localization length. It is ultimately due to the dispersion of the atom positions around the center of mass of the soliton.
Whether the localization or the delocalization effect wins depends on the parameter values. For the parameters used here, if $k\xi>1.8,$ the
localization length of the soliton is longer than the single atom localization length, shorter otherwise. 
Thus, no general statement
on whether attractive interactions favor or not Anderson localization can be made.

\section{Summary} 

To summarize, we have shown the existence of many-body Anderson localization for attractive bosons
in the presence of a disordered potential. The claim is based on quasi-exact many-body numerical simulations
using the TEBD algorithm, which incorporate all complicated phenomena that could spoil
the internal phase coherence of the many-body composite object, a bright soliton, displaying Anderson localization.
Moreover, we obtain excellent agreement between the many-body calculation and a one-body 
effective theory, which goes beyond
standard mean field theories such as the Gross-Pitaevskii equation.
Our quasi-exact many-body approach allows for simulation of the entire experiment 
starting from the initial state in a harmonic trap till the destructive measurement of 
all atom positions.

\section{Ancknowledgements}
Computing resources have been provided by GENCI and IFRAF. This work was performed within Polish-French bilateral
programme POLONIUM No.27742UE. Support of Polish National Science Center via projects DEC-2011/01/N/ST2/00418 (MP) and
 DEC-2012/04/A/ST2/00088 (KS and JZ) is acknowledged.

\end{document}